\documentclass[11pt]{article}

\usepackage[left=2.7cm,top=2cm,right=2.7cm,nohead]{geometry}
\usepackage{url}
\usepackage{color}
\usepackage[table]{xcolor}
\usepackage{graphicx}

\usepackage{etoolbox}
\BeforeBeginEnvironment{minted}{\begin{tcolorbox}[top=3pt,bottom=3pt]}%
\AfterEndEnvironment{minted}{\end{tcolorbox}}%

\usepackage[T1]{fontenc}
\usepackage{upquote}

\robustify{\texttt}
\let\originaltexttt\texttt

\begingroup
\catcode`'=\active
\catcode``=\active
\makeatletter
\renewrobustcmd{\texttt}[1]{%
   {%
   \everyeof{\noexpand}\endlinechar-1
   \expandafter\catcode\string``=\active
   \expandafter\catcode\string`'=\active
   \let'\textquotesingle
   \let`\textasciigrave
   \ifx\encodingdefault\upquote@OTone
    \ifx\ttdefault\upquote@cmtt
     \def'{\char13 }\def`{\char18 }%
    \fi
   \fi
   \scantokens{\originaltexttt{#1}}%
   }%
}%
\endgroup

\usepackage{adjustbox}
\newcolumntype{R}[2]{%
    >{\adjustbox{angle=#1,lap=\width-(#2)}\bgroup}%
    l%
    <{\egroup}%
}
\newcommand*\rot{\multicolumn{1}{R{60}{1em}}}

\usepackage{pifont}
\newcommand*\jau{\color{green!70!black}\ding{52}}
\newcommand*\mou{\color{red!60!white}\small\ding{56}}

\title{The Ocean Tensor Package}
\author{E. van den Berg\thanks{IBM T.J.~Watson Research Center, Yorktown
  Heights, NY, USA (\texttt{evandenberg@us.ibm.com})}}
\begin{document}

\maketitle

\begin{abstract}

  Matrix and tensor operations form the basis of a wide range of
  fields and applications, and in many cases constitute a substantial
  part of the overall computational complexity. The ability of
  general-purpose GPUs to speed up many of these operations and enable
  others has resulted in a widespread adaptation of these devices. In
  order for tensor operations to take full advantage of the
  computational power, specialized software is required, and currently
  there exist several packages (predominantly in the area of deep
  learning) that incorporate tensor operations on both CPU and
  GPU. Nevertheless, a stand-alone framework that supports general
  tensor operations is still missing. In this paper we fill this gap
  and propose the Ocean Tensor Library: a modular tensor-support
  package that is designed to serve as a foundational layer for
  applications that require dense tensor operations on a variety of
  device types. The API is carefully designed to be powerful,
  extensible, and at the same time easy to use. The package is
  available as open source.

\end{abstract}

\section{Introduction}

Over the last decade or so, general-purpose GPUs have been
successfully used in fields such as medical imaging
\cite{SHI2012LZXa,WAN2018PCLa}, molecular dynamics
simulations~\cite{KUT2015PFEa}, radio astronomy~\cite{BRO2018MNAa},
data mining~\cite{CAN2018a}, graph processing~\cite{SHI2018ZZJa}, and
many others~\cite{NVIDIA-2018-Applications}. More recently, GPUs have
found widespread use in, and, to a large extend, enabled deep
learning. Perhaps more so than in other fields, there has been a
proliferation of software packages for deep learning, such as
Caffe~\cite{JIA2014SDKa}, Torch~\cite{COL2011KFa}, and
Tensorflow~\cite{Tensorflow2015}. One reason for this is the
flexibility required in deep learning to try out different network
architectures and data transformations to obtain the best possible
model.
Given that the data, node parameters, and intermediate results are
conveniently expressed in the form of multi-dimensional arrays, or
tensors, these packages have seen a gradual shift towards
general-purpose compute environments. Despite these advances there is
still a lot of room for improvement: existing packages tend to be
monolithic, require a large number of external dependencies, and tend
to lack in tensor layout flexibility, supported data types, or
extensions to new device types. Most importantly, a stand-alone
tensor-support package designed to serve as the foundation for a wide
range of other applications is still missing. To fill this gap and
address some of the shortcomings of existing packages, we propose the
Ocean Tensor Package, a modular open-source\footnote{Available at
  \texttt{https://github.com/ibm/ocean-tensor-package}} foundation
library for dense tensor operations.

We describe the design and implementation of the Ocean Tensor Package,
or Ocean for short, in Section~\ref{Sec:Implementation} and highlight
some of its unique features in Section~\ref{Sec:Functionality}. In
Section~\ref{Sec:Existing} we take a look at existing packages and
contrast them against Ocean. We provide an illustrative example of using
Ocean in Section~\ref{Sec:Examples}, and conclude in
Section~\ref{Sec:Conclusions}.

\section{Design and implementation}\label{Sec:Implementation}\label{Sec:Modules}

The Ocean Tensor Package is designed to serve as a foundational layer
for applications that require dense tensor operations on one or more
device types and instances. Given the wide range of potential
applications and domains, it is important that the tensor operations
are grouped in coherent modules, rather than be provided through a
huge monolithic package. This way, functionality can be installed by
the user when needed, which helps reduce the effective number of
dependencies. Another advantage is that interfaces and compatibility
with external libraries is localized to independent modules, thus
making the package easier to manage. Another design principle used in
Ocean, and discussed later in this section, is the use of well-defined
layers.

\subsection{Modularity}

Modules in Ocean consist of an interface along with independent
implementations for each of the supported device types. The module
interface takes care of device-independent parameter checks including
validity of the tensors and compatibility of tensor dimensions. It
then determines the data type and device to use, and queries a
function look-up table for the module associated with the device
type. When available, the function is called after performing all
necessary type conversions, broadcasting, and allocation of result and
intermediate tensors (for instance when tensors overlap in memory). In
case the function is not available, or the module implementation for
the device type has not been loaded, an error is raised. Functions at
the device level typically only need to check for support of tensors
of the given data type and either implement the tensor operation or,
more typically, call a lower-level library function that provides the
desired operation. If needed, functions can access the module-specific
context information associated with each device instance.

Module interfaces and device implementations can be loaded separately,
except for the core module interface, which includes the CPU
implementation. The separation between the interface and the
implementation makes it possible to replace the module implementation
with alternatives, such as a highly-tuned or specialized proprietary
version. The use of function tables also makes it possible to replace
individual functions for performance comparisons or debugging, or to
insert functions that record runtime or accumulate call statistics.
The separation between module interfaces and device implementation
also make it easy to extend Ocean with new device types. In
particular, modules and functions within each module can be added and
tested one at a time, thus avoiding a huge development effort to get
started.

The Core module forms the basis of the Ocean Tensor Package. It
provides all elementary tensor operations and instantiates and exposes
available device instances. Many of the standard functions, such as
printing and tensor copies between different device types require
tensor support on the CPU, and the Core module interface is therefore
combined with the CPU implementation (\texttt{pyOcean\_cpu}). The GPU
implementation can be loaded separately by importing
\texttt{pyOcean\_gpu}. For convenience both packages are loaded by
importing \texttt{ocean}. Dependencies between modules are allowed, and
instantiation of one module can cause other modules to be loaded. The
instantiation order of modules is carefully registered to ensure that
modules are finalized only when no other modules depend on them.

\subsection{Layered implementation}

\begin{figure}
\centering
\includegraphics[width=0.82\textwidth]{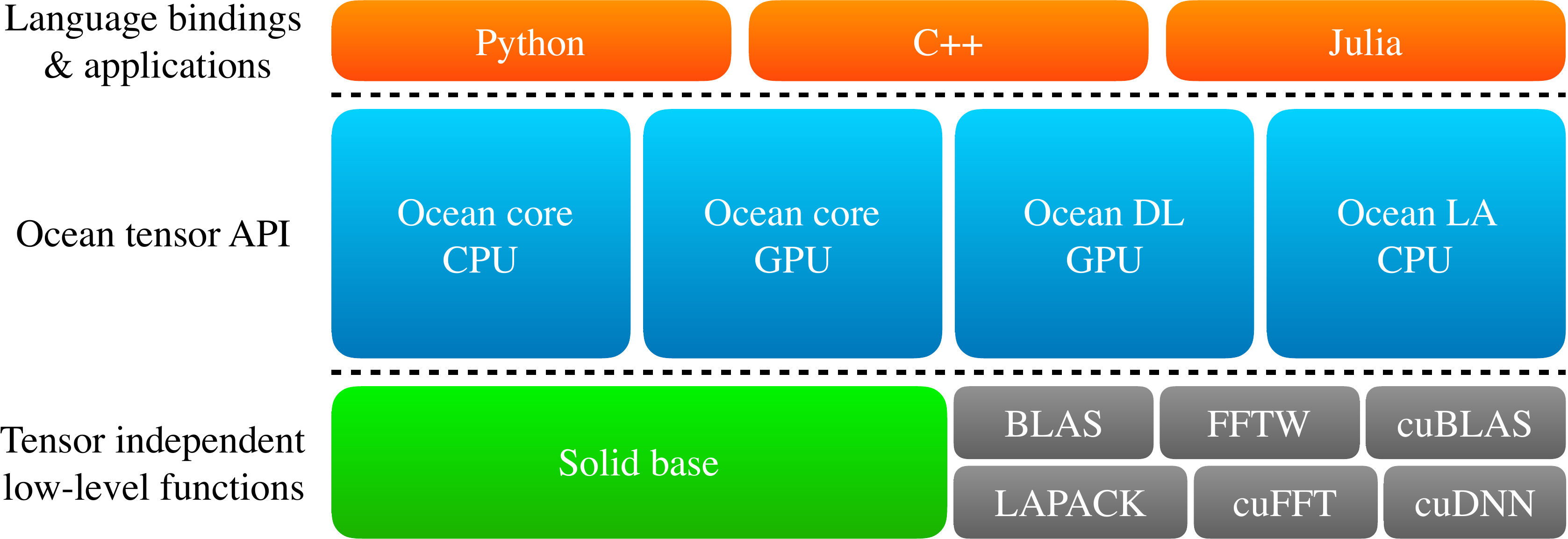}
\caption{Layered design of the Ocean Tensor Package. The current
  version implements the core modules for CPU and GPU, based on BLAS,
  cuBLAS, and the Solid foundation library, along with the Python
  language binding.}\label{Fig:Layers}
\end{figure}

In the implementation of the Ocean Tensor Package care is taken to
maintain a clean separation between the different abstraction levels,
as illustrated in Figure~\ref{Fig:Layers}. The libraries at the bottom
level provide low-level tensor operations that are independent on the
tensor representation. This includes existing libraries such as BLAS
and cuBLAS, as well as the custom developed Solid foundation library,
which provides elementary tensor functions for CPU and GPU. Libraries
at this level are not specific to Ocean and can be used independently
by other applications that require low-level tensor operations.
The Ocean tensor API defines a unified tensor type along with a
variety of tensor operations, organized as modules. As described
above, modules themselves can be classified in two layers, namely
interfaces and device implementations. The Ocean API is implemented in
C to maximize the accessibility from other languages.
The top level in Figure~\ref{Fig:Layers} shows possible language
bindings to Ocean (the current version only implements support for
Python). However, usage of the API is not restricted to language
bindings. For instance, applications that use tensor operations, or
libraries that support symbolic tensor compute graphs, too could be
build on top of the Ocean tensor API.

\section{Functionality}\label{Sec:Functionality}

The Ocean Tensor Package~\cite{Ocean2018} provides a comprehensive set
of tensor operations for CPU and GPU. The functions are available
directly as a C library, or through an easy-to-use Python interface.
In this section we explore some of the features of the package,
illustrated with code excerpts based on the Python interface.

\subsection{Object types}

\begin{figure}
\centering
\includegraphics[width=0.8\textwidth]{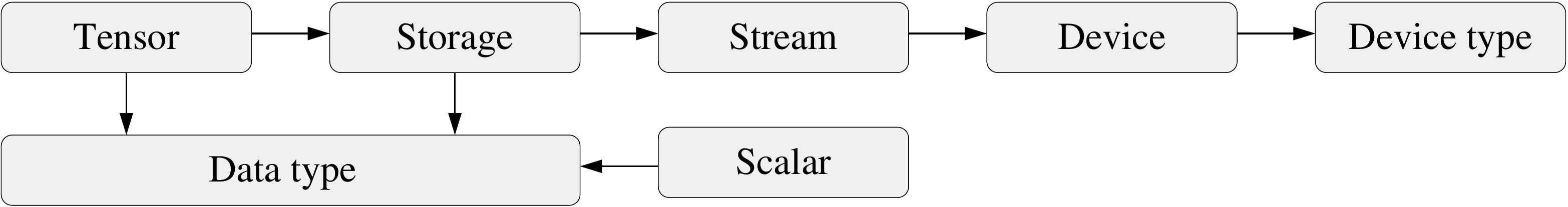}
\caption{Connection between the main object types in Ocean.}\label{Fig:ObjectTypes}
\end{figure}

The user interface to the Ocean Tensor Package exposes several
object types, illustrated in Figure~\ref{Fig:ObjectTypes}. At the top of
the object hierarchy are Tensors and Scalars, each of which have a
given data type. Tensors are views on contiguous chunks of memory,
stored as Storage objects. Associated with each Storage object is a
Stream, which is used to schedule asynchronous operations as well as
to maintain inter-stream dependencies. Stream object themselves are
associated with a Device instance (such as CPU or GPU \#0) of a certain
Device type (such as CPU or GPU).

\subsubsection{Devices}

Device objects enable the specification of the device to be used when
instantiating a Tensor or Storage object. In addition, they provide
generic information of the given device, such as the support for
byte-swapped data or the list of all currently loaded
modules. Depending on the device type, addition information may be
available. For instance, on GPU devices it is possible to query
numerous properties, including the multiprocessor count, or the
currently available free memory. Advanced functions include the
instantiation of a new stream, and the specification of the number of
intermediate tensor buffers for the device and their maximum
size. Ocean maintains a list of available devices, including the
\texttt{ocean.cpu} device as well as the \texttt{ocean.gpu} list of
GPU devices, which can be indexed to obtain the desired device
instance.

\subsubsection{Storage}

Storage objects encapsulate a contiguous piece of memory that is
either allocated dynamically or provided by and external source.  The
data type associated with the storage has two main purposes: first as
the default data type when instantiating a tensor from the storage
without providing a type; and second, for formatting the storage
elements for display. The data type of storage objects can be freely
changed without affecting the tensors that use it.
 It is possible to superimpose tensors of different data types on
the same storage. One typical example where this happens is when
querying the imaginary part of a complex double-precision tensor,
which results in an additional tensor view on the storage of type double.
There are no restrictions on the number of different tensor types that
can share the same storage. Tensor operations use the storage stream
for synchronization to avoid race conditions and inconsistencies in
the data. Storage data can be marked as read only, which prevents any
updates to the data, either directly or through tensors operations
(marking storage as read-only is reflected in all derived tensors).

\subsubsection{Tensors}

Ocean tensors are easy to instantiate on any of the available
devices. For example, creation of a $3\times 4$ tensor with
single-precision floating-point format on device \texttt{gpu[0]} is
done simply by writing:
\texttt{tensor = ocean.tensor([3,4], ocean.float, ocean.gpu[0])}.
When the data type or device is omitted, user-specified default values
are used. In contrast to some of the other packages, described later
in Section~\ref{Sec:Existing}, there is no notion of a currently
active device, and no explicit device changes need to be done in order
to instantiate new tensors, or perform operations on them. By default,
tensors follow a column-major memory layout, but general strides
(given in bytes) are supported, thereby allowing compatibility with
most Numpy tensors. Two differences with Numpy~\cite{OLI2006a} are (1)
the support of the complex half-precision data type in Ocean, and
additional data types in Numpy (such as string and datetime), and (2)
the maximum number of tensors dimensions. The maximum number of tensor
dimensions in Ocean is currently set to eight, but this restriction is
easy relaxed or removed; Numpy has a hard-coded maximum of 32 tensor
dimensions.  Similar to Numpy, Ocean allows tensors on the CPU to have
either little and big endian byte order, and tensor operations can be
applied in either byte order. The byte order can be easily changed, if
needed, either by byte-swapping the elements, or, in case of a
mismatch between the flag and the actual byte ordering, by simply
specifying the appropriate byte order.

\subsection{Tensor operations}

Tensor operations in Ocean are provided through modules, as explained
in Section~\ref{Sec:Modules}. The Core module forms the basis of
Ocean, and includes the definition of the basic object classes as well
as the device instances. As the most elementary operation, the Core
module supports tensor creation; with or without initialization, from
storage, as well as from data in the form of nested lists, sequences,
and other tensor types. Aside from tensor creation, the Core module
provides an extensive set of elementary functions, including functions
for shape and axis manipulations, indexing, copy functions, type and
device casting, basic arithmetic operations, trigonometric operations
(supported on all real and complex floating-point types), as well as
tensor reductions along one or more axes. (A complete list of
functions can be found on the Ocean Tensor Package
repository~\cite{Ocean2018}.) Below we highlight some of the
functionalities.

\subsubsection{Type casting}\label{Sec:TypeCasting}

The type of a tensor can be seen as the combination of the data type
and the device associated with the tensor. Tensors in Ocean have an
associated type, and type casting may therefore be necessary at
times. Explicit type casting can be using the \texttt{ocean.cast}
function, which returns a copy of the tensor with the desired type,
and the \texttt{ocean.ensure} function, which returns a type-cast
tensor only if the requested type differs from that of the input. Type
casting with a data type (\texttt{ocean.float(T)}) or device instance
(\texttt{ocean.gpu[1](T)}) is equivalent to calling the ensure
function with only a data type or device update.

Implicit type casting in Ocean is used to ensure that the input
arguments to tensor operations have appropriate types and byte
order. Consider for instance the tensor addition: \texttt{C = A+B}. To
avoid implementing addition for all possible type combinations we need
to determine the type of \texttt{C} based on those of \texttt{A} and
\texttt{B}, and normalize the data types and device accordingly. One
way of resolving the device type is to impose a device ordering and
choose the device with the highest priority. This requires
specification of the order and may result in unexpected results,
certainly when the device order could be changed from other parts of
the code. Another way is to always use the device from the left-hand
side of the operator, or the first parameter in the argument list. In
\texttt{A+=B} it is clear that \texttt{B} should be coerced to the
device of \texttt{A}, and we therefore do the same for
\texttt{A+B}. In case it is desirable to use the device of \texttt{B}
(i.e., \texttt{B.device}), it possible in this case to write
\texttt{B+A}, or to use explicit casting:
\texttt{ocean.ensure(A,B.device) + B}, or simply \texttt{B.device(A) +
  B}.

For implicit casting of the data types we follow Numpy and use the
smallest available data type that can keep both data types. For
instance, addition of signed and unsigned 8-bit integers would give a
16-bit signed integer. Since no standard data type is available for
quadruple-precision floats, an exception is made for 64-bit integers
and floating-point numbers, which result in double-precision floats.
Automatic type casting on Ocean is switched on by default, but can be
disabled by the user in case strict type checks are needed. When
switched off, an exception is raised whenever a type mismatch is
encountered.

Type casting based on the contents of the tensor is desirable, for
example, when taking the square root of negative real numbers or the
arccosine of scalars with magnitude larger than one. Should such
operations result in a not-a-number (NaN) value, return a
complex-valued result, or generate an error? The approach taken in
Ocean is to add parameters to such operators that indicate the compute
mode. In the standard mode no checks on the tensor elements are done
and NaN values are generated whenever needed. Checks are made in the
warning and error modes, giving respectively a warning or an error
when elements with values outside of the operator domain are
encountered. Finally, in the complex mode, checks are made to
determine whether the resulting data type should be real or
complex. If needed, explicit types casting can always be used.

\subsubsection{Indexing}

Ocean supports a variety of indexing modes along one or more
dimensions that can be combined to index a tensor. The basic
single-dimension indexing modes are (1) scalars, to index a single
element along an axis; (2) ranges, to index a regularly spaced set of
elements; and (3) the colon `\texttt{:}' operator to indicate the
entire dimension. In addition to the basic modes it is possible to use
one or two-dimensional index arrays to select particular elements, by
specifying the indices along a single dimension, or tuples of indices
along several dimensions. As is customary in Python, negative indices
can be used to indicate the index relative to the end of the
dimension. Finally, boolean tensors can be used as masks for indexing,
with the non-zero elements indicating the elements to be
selected. Dimensions that are omitted in the indexing are implicitly
indexed with the colon operator, and the ellipsis object
`\texttt{...}' can appear once to indicate application of zero or more
colon operator in that location to complete the index. When indexing a
tensor using only basic indexing modes (either explicitly or
implicitly), a view of that tensor is returned in the form of a new
tensor that shares the original storage. In all other cases a new
tensor is created by copying the indexed elements.

Special preprocessing is needed for index arrays and boolean masks:
for index arrays the indices need to be checked for validity; whereas
for boolean masks it is necessary to count the number of selected
elements, in order to determine the size of the output tensor; and to
convert the selected indices into relative offsets into the data
buffer of tensor being indexed. When such indices are used repeatedly,
computational efforts are wasted in applying the same preprocessing
steps for each use. To avoid this situation, Ocean introduces index
objects, which are constructed by indexing the \texttt{ocean.index}
object (ranges and colons parameters are not allowed as parameters in
regular function calls). Once an index object is created is can be
bound to a tensor size to convert negative indices, check the validity
of indices, determine the overlap of index ranges with the given
dimensions, and convert boolean masks to explicit indices. Index
objects can subsequently (or directly) be bound to tensor strides,
which converts all index arrays and boolean masks to relative offsets
within the tensor data. Both bound and unbound index object can be
used in exactly the same way as the index modes used to construct
it. As such, index object can be used to create yet other index
objects, if needed. When index objects are bound to size, the relevant
tensor dimensions must match, and likewise for strides.

\subsubsection{Interoperability}

The Python interface of Ocean provides for plug-in modules to define
external object types for tensors and scalars, and the conversion
between these and the corresponding Ocean types. All extended object
types provided by the plug-ins are compared against when parsing the
tensor operation parameters. This allows them to be used in
essentially the same way as Ocean tensors and scalars. As an example,
we can declare the Numpy tensor and scalar types by importing
\texttt{pyOceanNumpy}. Once this is done, it is possible to write
expressions such as \texttt{A + np.asarray([4,5,6])}, where \texttt{A}
is an Ocean tensor. Conversion to Numpy can be done using
\texttt{A.convertTo('numpy')}, where the \texttt{'numpy'} string is
registered by the plug-in. When supported by the external tensor type,
a shallow copy of the tensor is made, unless otherwise requested by
the user.

\subsubsection{Deallocation}

Automatic garbage collection in Python can delay the deletion of
tensor objects and cause devices to run out of free memory despite
careful management by the user. In order to force tensor deletion, it
is possible to call the \texttt{dealloc} tensor function, which
maintains the Python tensor object, but replaces the content by an
empty tensor. This frees up any dynamically allocated tensor data,
while avoiding problems with accidental usage of the tensor after
deallocation.

\section{Existing packages}\label{Sec:Existing}

We now compare some of the features in Ocean with those in other
packages. Since most of the packages are in active development, we
only discuss the features available at the time of writing.

Numpy~\cite{OLI2006a} is the {\it{de facto}} Python package for dense
tensor operations. Numpy was written for tensors on CPUs and does not
support tensors on any other device types. The more recent
CuPy~\cite{OKU2017UNHa} package implements tensors for GPU devices
with an interface that closely mirrors that of Numpy, but is otherwise
largely independent. Both packages are written as a Python-C API and
directly extend Python classes, which limits their usage as
stand-alone packages. Moreover, each of these two package supports
only a single device type.

\begin{table}[t]
\centering
\begin{tabular}{lccccccccc}
&\rot{Numpy}
&\rot{CuPy}
&\rot{Caffe}
&\rot{PyTorch}
&\rot{TensorFlow}
&\rot{MXNet}
&\rot{ArrayFire}
&\rot{Ocean} \\
\hline
Multiple device types & \mou & \mou & \jau & \jau & \jau & \jau & \jau & \jau \\
Automatic type casting & \jau & \jau & \mou & \mou & \mou & \mou & \jau & \jau \\
Unified tensor type & \jau & \jau & \mou & \jau & \jau & \jau & \jau & \jau \\
Complex data types & \jau & \jau & \mou & \mou & \jau & \mou & \jau & \jau \\
Flexible tensor strides & \jau & \jau & \mou & \jau & \mou & \mou &
                                                                    \mou & \jau\\
Tensor overlap detection & \jau & \mou & \mou & \jau & \jau & \mou & \mou & \jau \\
\hline
\end{tabular}
\caption{Comparison between different packages providing tensor functionality.}\label{Table:Comparison}
\end{table}

A package that supports multiple device types and is written as a
general library with separate language bindings is
ArrayFire~\cite{YAL2015AMGa}. The same applies to most deep-learning
packages. As mentioned in the introduction, tensor operations form the
foundation of deep-learning packages, and we therefore consider some
of the most popular ones: Caffe~\cite{JIA2014SDKa},
PyTorch~\cite{COL2002BMa, PAS2017GCCa},
TensorFlow~\cite{Tensorflow2015}, and MXNet~\cite{CHE2015LLLa}. All of
these packages support tensor operations on multiple device types, and
expose at least part of the available tensor operations to the
user. Nevertheless, given the focus on deep learning, these packages
are not written or intended to serve as stand-alone tensor support
packages. In particular, many of the packages define tensor classes
with highly domain-specific member functions and variables. For
example, the class may provide gradient information with each tensor,
or include references to a symbolic compute graph node that contains
the tensor.

In Table~\ref{Table:Comparison} we list several properties that we
consider to be important in a general-purpose tensor package, and are
therefore implemented in Ocean. One of these properties is the support
for automatic type casting, as discussed in detail in
Section~\ref{Sec:TypeCasting}. This feature is supported in Numpy,
CuPy, and ArrayFire, but is missing from all the deep-learning
packages under consideration.  From a developer's point of view it is
convenient to have a unified tensor type or class. This is supported
by all packages except Caffe, which provides a templated class
(several of the other packages use templated classes behind the
scenes, but provide a unified type in the API).  Support for a
comprehensive set of data types is clearly important for a tensor
package\footnote{Ocean supports booleans; 8, 16, 32, and 64-bit signed
  and unsigned integers; as well as real and complex half, single and
  double-precision floating-point data types.}. Coverage between
packages differs substantially, and we therefore focus on support for
complex data types, which required additional functionality, such as
conjugation and access to real and imaginary parts of the tensor. Of
the four deep-learning packages considered, only TensorFlow supports
complex tensor types (based on single and double-precision
floats). These types are also supported by Numpy, CuPy, and ArrayFire.
Ocean is the only package that additionally provides a complex data
type based on half-precision floats.

The layout of tensors in memory is given by the strides, or distance
between successive elements along each of the dimensions. Flexibility
in the tensor strides enables features such as broadcasting along
dimensions, easy manipulation of axes, and creation of views on
regularly indexed subtensors. In addition, it ensures compatibility
with a wide range of existing data types for tensors and
matrices. Most of the deep-learning packages, as well as ArrayFire,
adhere to a contiguous row-major data order, with implicit strides
that can be inferred based on the tensor dimensions and the element
size of the given data type. PyTorch also uses this data order by
default, but allows users to override the standard layout by
specifying tensor strides as nonnegative multiples of the element
size. Numpy and CuPy support arbitrary strides. Each of these
packages, along with Ocean, implements sorting of axes and merging of
consecutive axes when possible, to increase memory locality and reduce
the overhead of iterating over dimensions, both of which help increase
the computational efficiency of tensor operations on strided data.
Packages that use a contiguous tensor layout can flatten tensors to a
single dimension for many operations, such as unary and binary
elementwise operations; other operations may require optimizations
similar to those mentioned above. ArrayFire limits the number of
tensor dimensions to four and often uses explicit nested for-loops
with index computation at the innermost loop to traverse the data.

Some of the difficulties that come with the provision for arbitrary
strides is that tensors may self overlap in memory, and that overlap
detection between pairs of tensors becomes non-trivial. For consistent
results in computations, such as $A[[1,2]] = A[[2,1]]$, good support
for overlap detection is essential. Ocean checks for self-overlapping
tensors and treats them as read-only for most operations (the
semantics of writing different values to the same memory address is
not well defined). Overlap detection between pairs of tensors and
allocation of intermediate tensors to resolve overlaps is also
included. Similar checks are present in PyTorch and Numpy. Overlap
detection in TensorFlow is restricted to tensor views.

Aside from Ocean, none of the packages considered in this section
defines a clear separation between tensor types and the low-level
implementation of the tensor operations. As a result, none of the
tensor operations other than those already provided by existing
libraries, such as BLAS and cuBLAS, are easily transferable for use in
other packages.

\section{Illustrative example}\label{Sec:Examples}

\begin{figure}
\centering
\footnotesize
\begin{tabular}{cc}
\includegraphics[width=0.39\textwidth]{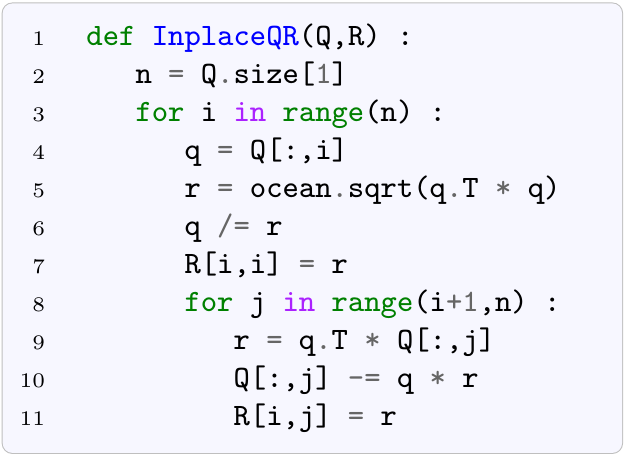}&
\includegraphics[width=0.37\textwidth]{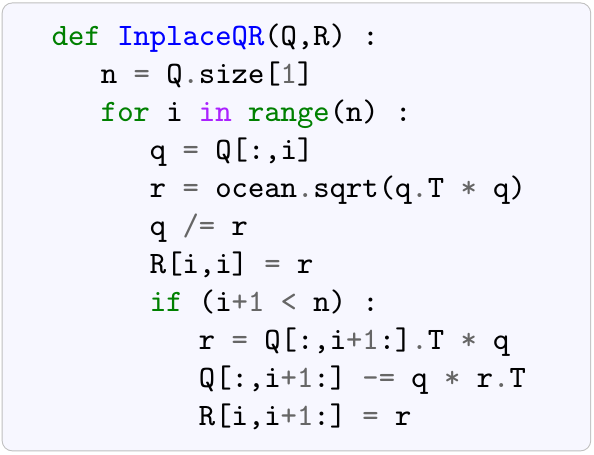}\\
{\footnotesize({\bf{a}})} &
{\footnotesize({\bf{b}})}
\end{tabular}\\[5pt]

\scriptsize
\includegraphics[width=0.784\textwidth]{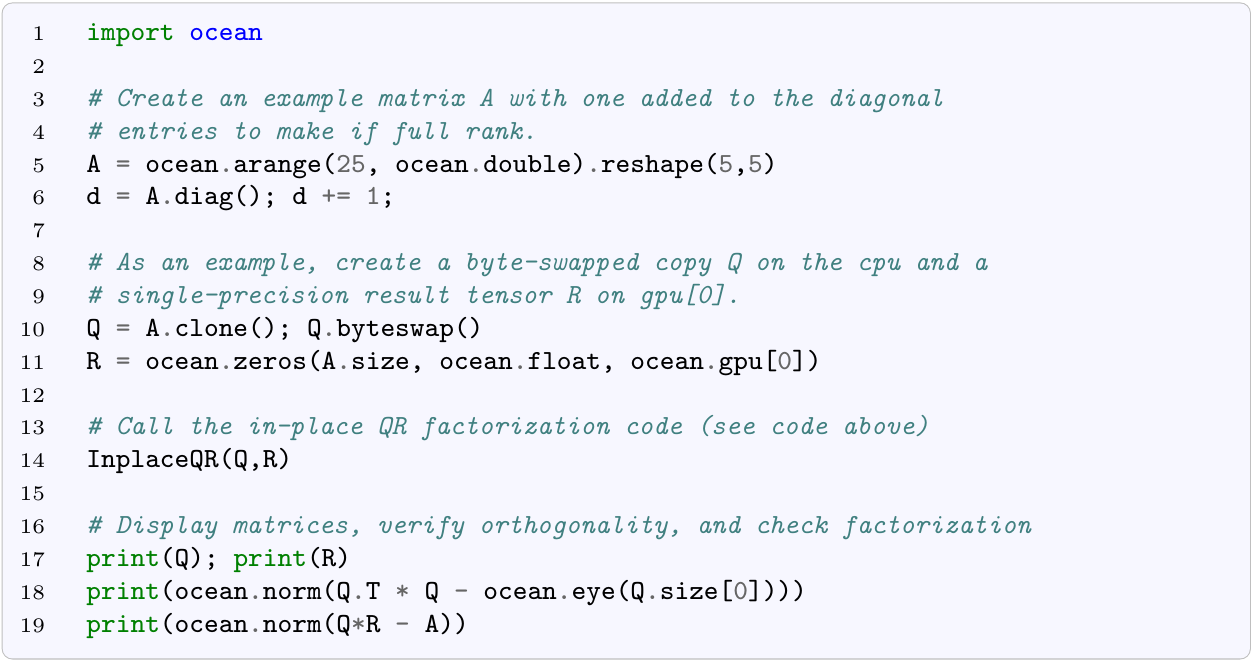}\\
{\footnotesize({\bf{c}})}\\[5pt]

\scriptsize
\includegraphics[width=0.784\textwidth]{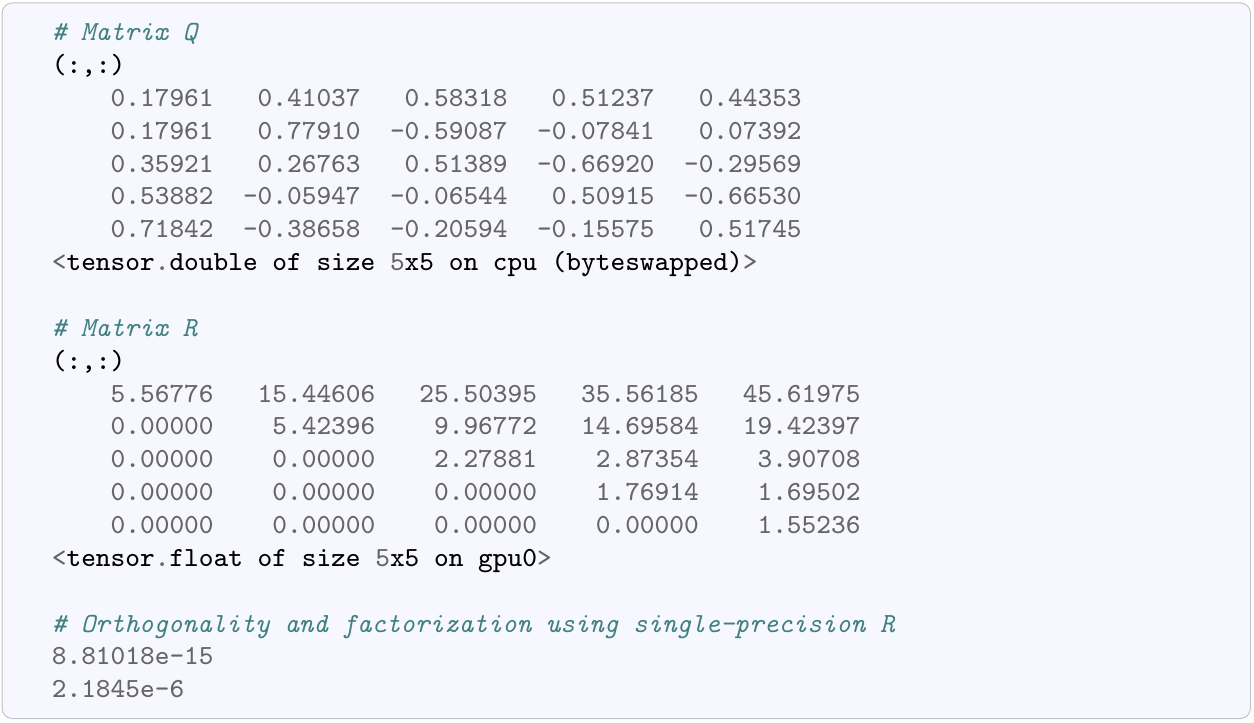}\\
{\footnotesize({\bf{d}})}
\caption{In-place QR factorization functions with (a) individual
  column updates in $Q$, and (b) multi-column rank-one matrix updates
  of $Q$; along with (c) the calling script, and (d) the output
  annotated with comments. Note that Ocean supports vector inner and
  outer products of the form \texttt{u.T*v} and \texttt{u*v.T}.}\label{Fig:QR}
\end{figure}

We now illustrate some of the features of the Ocean Tensor Package
based on an example QR factorization (see for
instance~\cite{TRE1997Ba}). This is of course only an example; QR
factorization should be an integral part of the package, and direct
support is planned in a future linear-algebra module. A comprehensive
list of functions provided by the core module, as well as a large
number of examples can be found in the documentation of the Ocean
Tensor Package repository~\cite{Ocean2018}.

The functions shown in Figures~\ref{Fig:QR}(a) and~\ref{Fig:QR}(b)
implement two variations of an in-place QR factorization $A = QR$. The
original matrix $A$ is stored in the first parameter, $Q$, and will be
overwritten by the orthonormal part of the factorization. The second
parameter, $R$, is assumed to be pre-allocated and will be used to
store the upper-diagonal part of the factorization. Normally the only
input to the function would be $A$, but here we allow the user to
specify $Q$ and $R$ with any data type and device. In line 2 of the
code we determine the number of columns in $Q$. The QR factorization
then proceeds by repeating the following process on each of the
columns $i$ in $Q$. First we get the current column \texttt{q =
  Q[:,i]} in line 4. Because the indexing is regular this tensor will
be a view of the original $Q$. In line 5 we determine the two-norm of
$q$ by taking the square root of the inner-product with itself. Note
here, that unlike many other packages we define the transpose
\texttt{q.T} as a $1\times n$ matrix rather than as the original
one-dimensional vector. We normalize $q$, and therefore the
corresponding column in $Q$, in line 6, and update the diagonal entry
in $R$. The for-loop in lines 8--11 in Figure~\ref{Fig:QR}(a) projects
out the $q$ component in each of the remaining columns of $Q$,
proceeding one column at a time. The same lines in
Figure~\ref{Fig:QR}(b) achieve the same, operating on all remaining
columns at once using indexing and matrix operations instead of vector
operations. In the second case we take the outer product of vectors
$q$ and $r$, written as $qr^T$.

Figure~\ref{Fig:QR}(c) gives an example script calling either one of
the two factorization functions (algorithmically they are equivalent).
After importing the Ocean Tensor Package in line 1 we create a square
matrix $A$ on line 5, by first creating a vector of type double on the
default device (CPU) containing the numbers 0 through 24 of type
double, and then reshaping the vector to a $5\times 5$ matrix. The
resulting matrix has rank 2 instead of 5, and in line 6 we therefore
add the identity matrix by creating a view $d$ on the diagonal and
adding 1 to each element. On lines 10--11 we prepare $Q$ by creating a
deep copy of matrix $A$ and byteswapping its data, and allocate a
float matrix $R$ on device \texttt{gpu[0]} with the same dimensions as
$A$. We choose $Q$ and $R$ to have different data type, device, and
byte order to illustrate the seamless type casting provided by Ocean.
After calling the in-place QR factorization function in line 14, we
print out the two matrices along with the Frobenius norms
$\Vert Q^TQ - I\Vert_F$ and $\Vert QR-A\Vert_F$ in lines 17--19. The
output of the script given in Figure~\ref{Fig:QR}(d) shows the
elements and type of matrices $Q$ and $R$, and the values of the two
norms. The seemingly inaccurate factorization of $A$ is due
to the use of single-precision floats in $R$.

\section{Conclusions}\label{Sec:Conclusions}

In this paper we presented the Ocean Tensor Package, a general-purpose
tensor support package for dense tensors on different device
types. The package is organized in a modular fashion, with coherent
sets of tensor operations grouped together in modules. Each module
consists of a device-independent interface that exposes the available
functions, along with separate device-specific modules that provide
the implementation of these functions on these devices. A conscious
design decision in Ocean was to provide clearly separated layers of
abstraction. The bottom layer provides low-level tensor operations
that are independent of the tensor representation and can therefore
also be used as a stand-alone foundation library by other
packages. The foundation of the Ocean Tensor Package is firmly
established in the core module, and future work will therefore largely
focus on increasing the functionality through the addition of new
modules and expansion of existing ones.

\bibliography{bibliography}

\begin{thebibliography}{10}

\bibitem{Tensorflow2015}
M.~Abadi, A.~Agarwal, P.~Barham, E.~Brevdo, Z.~Chen, C.~Citro, G.~S. Corrado,
  A.~Davis, J.~Dean, M.~Devin, S.~Ghemawat, I.~Goodfellow, A.~Harp, G.~Irving,
  M.~Isard, Y.~Jia, R.~Jozefowicz, L.~Kaiser, M.~Kudlur, J.~Levenberg,
  D.~Man\'{e}, R.~Monga, S.~Moore, D.~Murray, C.~Olah, M.~Schuster, J.~Shlens,
  B.~Steiner, I.~Sutskever, K.~Talwar, P.~Tucker, V.~Vanhoucke, V.~Vasudevan,
  F.~Vi\'{e}gas, O.~Vinyals, P.~Warden, M.~Wattenberg, M.~Wicke, Y.~Yu, and
  X.~Zheng.
\newblock {TensorFlow}: Large-scale machine learning on heterogeneous systems,
  2015.
\newblock Software available from tensorflow.org.

\bibitem{Ocean2018}
E.~van~den Berg.
\newblock The {O}cean {T}ensor {P}ackage, 2018.
\newblock \\Available at \url{https://github.com/ibm/ocean-tensor-package}.

\bibitem{BRO2018MNAa}
P.~C. Broekema, J.~J.~D. Mol, R.~Nijboer, A.~S. van Amesfoort, M.~A. Brentjens,
  G.~M. Loose, W.~F.~A. Klijn, and J.~W. Romein.
\newblock A {GPU}-based correlator and beamformer for {LOFAR}.
\newblock arXiv:1801.04834, 2018.

\bibitem{CAN2018a}
A.~Cano.
\newblock A survey on graphic processing unit computating for large-scale data
  mining.
\newblock {\em {WIREs} Data Mining Knowledge discovery}, 8:e1232, 2018.

\bibitem{CHE2015LLLa}
T.~Chen, M.~Li, Y.~Li, M.~Lin, N.~Wang, M.~Wang, T.~Xiao, B.~Xu, C.~Zhang, and
  Z.~Zhang.
\newblock {MXNet}: A flexible and efficient machine learning library for
  heterogeneous distributed systems.
\newblock In {\em Neural Information Processing Systems, Workshop on Machine
  Learning Systems}, 2015.

\bibitem{COL2002BMa}
R.~Collobert, S.~Bengio, and J.~Marithoz.
\newblock Torch: A modular machine learning software library, 2002.

\bibitem{COL2011KFa}
R.~Collobert, K.~Kavukcuoglu, and C.~Farabet.
\newblock Torch7: A {M}atlab-like environment for machine learning.
\newblock In {\em BigLearn, NIPS Workshop, number EPFL-CONF-192376}, 2011.

\bibitem{JIA2014SDKa}
Y.~Jia, E.~Shelhamer, J.~Donahue, S.~Karayev, J.~Long, R.~Girshick,
  S.~Guadarrama, and T.~Darrell.
\newblock Caffe: Convolutional architecture for fast feature embedding.
\newblock In {\em Proceedings of the 22nd ACM International Conference on
  Multimedia}, pages 675–--678, 2014.

\bibitem{KUT2015PFEa}
C.~Kutzner, S.~P\'{a}ll, M.~Fechner, A.~Esztermann, B.~L. de~Groot, and
  H.~Grubm\"{u}ller.
\newblock Best bang for your buck: {GPU} nodes for {GROMACS} biomolecular
  simulations.
\newblock {\em Journal of Computational Chemistry}, 36(26):1990--2008, 2015.

\bibitem{NVIDIA-2018-Applications}
{NVIDIA}.
\newblock {GPU}-accelerated applications.
\newblock
  \url{https://www.nvidia.com/en-us/data-center/gpu-accelerated-applications/catalog/}.

\bibitem{OKU2017UNHa}
R.~Okuta, Y.~Unno, D.~Nishino, S.~Hido, and C.~Loomis.
\newblock {CuPy}: A {NumPy}-compatible library for {NVIDIA} {GPU} calculations.
\newblock In {\em Proceedings of Workshop on Machine Learning Systems
  (LearningSys) in The Thirty-first Annual Conference on Neural Information
  Processing Systems (NIPS)}, 2017.

\bibitem{OLI2006a}
T.~E. Oliphant.
\newblock {\em A guide to {Numpy}}.
\newblock Trelgol Publishing, 2006.

\bibitem{PAS2017GCCa}
A.~Paszke, S.~Gross, S.~Chintala, G.~Chanan, E.~Yang, Z.~{DeVito}, Z.~Lin,
  A.~Desmaison, L.~Antiga, and A.~Lerer.
\newblock Automatic differentiation in {PyTorch}.
\newblock In {\em NIPS 2017 Autodiff Workshop}, 2017.

\bibitem{SHI2012LZXa}
L.~Shi, W.~Liu, H.~Zhang, Y.~Xie, and D.~Wang.
\newblock A survey of {GPU}-based medical image computing techniques.
\newblock {\em Quantitative Imaging in Medicine and Surgery}, 2(3):188--206,
  2012.

\bibitem{SHI2018ZZJa}
X.~Shi, Z.~Zheng, Y.~Zhou, H.~Jin, L.~He, B.~Liu, and Q.-S. Hua.
\newblock Graph processing on {GPUs}: A survey.
\newblock {\em ACM Computing Surveys}, 50(6):81:1--35, 2018.

\bibitem{TRE1997Ba}
L.~N. Trefethen and D.~{Bau, III}.
\newblock {\em Numerical Linear Algebra}.
\newblock SIAM, 1997.

\bibitem{WAN2018PCLa}
H.~Wang, H.~Peng, Y.~Chang, and D.~Liang.
\newblock A survey of {GPU}-based acceleration techniques in mri
  reconstructions.
\newblock {\em Quantitative Imaging in Medicine and Surgery}, 8(2):196--208,
  2018.

\bibitem{YAL2015AMGa}
P.~Yalamanchili, U.~Arshad, Z.~Mohammed, P.~Garigipati, P.~Entschev,
  B.~Kloppenborg, J.~James, and J.~Melonakos.
\newblock {ArrayFire} - a high performance software library for parallel
  computing with an easy-to-use {API}, 2015.
\newblock \url{https://github.com/arrayfire/arrayfire}.

\end{thebibliography}

\end{document}